

\font\titlefont = cmr10 scaled\magstep 4
\font\sectionfont = cmr10
\font\littlefont = cmr5
\font\eightrm = cmr8 

\def\sss{\scriptscriptstyle} 

\magnification = 1200

\global\baselineskip = 1.2\baselineskip
\global\parskip = 4pt plus 0.3pt
\global\abovedisplayskip = 18pt plus3pt minus9pt
\global\belowdisplayskip = 18pt plus3pt minus9pt
\global\abovedisplayshortskip = 6pt plus3pt
\global\belowdisplayshortskip = 6pt plus3pt


\def\endignore{}
\def\ignore #1\endignore{}

\newcount\dflag
\dflag = 0


\def\monthname{\ifcase\month
\or Jan \or Feb \or Mar \or Apr \or May \or June%
\or July \or Aug \or Sept \or Oct \or Nov \or Dec
\fi}

\def\timestring{{\count0 = \time%
\divide\count0 by 60%
\count2 = \count0
\count4 = \time%
\multiply\count0 by 60%
\advance\count4 by -\count0
\ifnum\count4 < 10 \toks1 = {0}
\else \toks1 = {} \fi%
\ifnum\count2 < 12 \toks0 = {a.m.}
\else \toks0 = {p.m.}
\advance\count2 by -12%
\fi%
\ifnum\count2 = 0 \count2 = 12 \fi
\number\count2 : \the\toks1 \number\count4%
\thinspace \the\toks0}}

\def\today{\ifcase\month\or January\or February\or March\or
 April\or May\or June\or July\or August\or September\or
 October\or November\or December\fi \space\number\day, \number\year}



\def\endtitle{}
\def\title#1\endtitle{\vskip.5in\titlefont
\global\baselineskip = 2\baselineskip
#1\vskip.4in
\baselineskip = 0.5\baselineskip\rm}
 
\def\endauthors{}
\def\authors#1\endauthors{#1}

\def\endabstract{}
\def\abstract#1\endabstract{\vskip .3in%
\centerline{\sectionfont\bf Abstract}%
\vskip .1in
\noindent#1}

\newcount\nsection
\newcount\nsubsection

\def\section#1{\global\advance\nsection by 1
\nsubsection=0
\bigskip\noindent\centerline{\sectionfont \bf \number\nsection.\ #1}
\bigskip\rm\nobreak}

\def\subsection#1{\global\advance\nsubsection by 1
\bigskip\noindent\sectionfont \sl \number\nsection.\number\nsubsection)\
#1\bigskip\rm\nobreak}

\def\topic#1{{\medskip\noindent $\bullet$ \it #1:}} 
\def\endtopic{\medskip}

\def\appendix#1#2{\bigskip\noindent%
\centerline{\sectionfont \bf Appendix #1.\ #2}
\bigskip\rm\nobreak}


\newcount\nref
\global\nref = 1

\def\ref#1#2{\xdef #1{[\number\nref]}
\ifnum\nref = 1\global\xdef\therefs{\noindent[\number\nref] #2\ }
\else
\global\xdef\oldrefs{\therefs}
\global\xdef\therefs{\oldrefs\vskip.1in\noindent[\number\nref] #2\ }%
\fi%
\global\advance\nref by 1
}

\def\listrefs{\vfill\eject\section{References}\therefs}


\newcount\nfoot
\global\nfoot = 1

\def\foot#1#2{\xdef #1{(\number\nfoot)}
\footnote{${}^{\number\nfoot}$}{\eightrm #2}
\global\advance\nfoot by 1
}


\newcount\nfig
\global\nfig = 1

\def\fig#1{\xdef #1{(\number\nfig)}
\global\advance\nfig by 1
}


\newcount\cflag
\newcount\nequation
\global\nequation = 1
\def\eqlabel{(1)}

\def\nexteqno{\ifnum\cflag = 0
\global\advance\nequation by 1
\fi
\global\cflag = 0
\xdef\eqlabel{(\number\nequation)}}

\def\lasteqno{\global\advance\nequation by -1
\xdef\eqlabel{(\number\nequation)}}

\def\label#1{\xdef #1{(\number\nequation)}
\ifnum\dflag = 1
{\escapechar = -1
\xdef\draftname{\littlefont\string#1}}
\fi}

\def\clabel#1#2{\xdef\eqlabel{(\number\nequation #2)}
\global\cflag = 1
\xdef #1{\eqlabel}
\ifnum\dflag = 1
{\escapechar = -1
\xdef\draftname{\string#1}}
\fi}

\def\cclabel#1#2{\xdef\eqlabel{#2)}
\global\cflag = 1
\xdef #1{\eqlabel}
\ifnum\dflag = 1
{\escapechar = -1
\xdef\draftname{\string#1}}
\fi}


\def\eeq{}

\def\eqnn #1\eeq{$$ #1 $$}

\def\eq #1\eeq{\xdef\draftname{\ }
$$ #1
\eqno{\eqlabel \rlap{\ \draftname}} $$
\nexteqno}



\def\eol{& \eqlabel \rlap{\ \draftname} \crcr
\nexteqno
\xdef\draftname{\ }}

\def\eeol{& \eqlabel \rlap{\ \draftname}
\nexteqno
\xdef\draftname{\ }}

\def\eolnn{\cr
\global\cflag = 0
\xdef\draftname{\ }}


\def\eqa #1\eeq{\xdef\draftname{\ }
$$ \eqalignno{ #1 } $$
\global\cflag = 0}


\def\etc{{\it etc.\/}}


\def\npb#1#2#3{{\it Nucl.~Phys.} {\bf B#1} (19#2) #3}
\def\plb#1#2#3{{\it Phys.~Lett.} {\bf #1B} (19#2) #3}

\def\prd#1#2#3{{\it Phys.~Rev.} {\bf D#1} (19#2) #3}

\def\prl#1#2#3{{\it Phys.~Rev.~Lett.} {\bf #1} (19#2) #3}


\global\nulldelimiterspace = 0pt



\def\frac#1#2{{{#1} \over {#2}}\,}  
\def\hf{{1\over 2}}



\def\Dsl{\hbox{/\kern-.6700em\it D}} 
\def\dsl{\hbox{/\kern-.5300em$\partial$}}
\def\pxpsl{\hbox{/\kern-.5600em$p$}}
\def\ssl{\hbox{/\kern-.5300em$s$}}
\def\epssl{\hbox{/\kern-.5100em$\epsilon$}}
\def\delsl{\hbox{/\kern-.6300em$\nabla$}}
\def\lxpsl{\hbox{/\kern-.4300em$l$}}
\def\elxpsl{\hbox{/\kern-.4500em$\ell$}}
\def\kxpsl{\hbox{/\kern-.5100em$k$}}
\def\qxpsl{\hbox{/\kern-.5000em$q$}}
\def\sla#1{\raise.15ex\hbox{$/$}\kern-.57em #1}



\def\roughly#1{\mathrel{\raise.3ex\hbox{$#1$\kern-.75em\lower1ex\hbox{$\sim$}}}}
\def\lsim{\roughly<}
\def\gsim{\roughly>}

\def\tw#1{\tilde{#1}}
\def\ol#1{\overline{#1}}





\def\Scz{{\cal Z}}






\def\sm{standard model}


\overfullrule=0pt
 

\def\dpi{\delta \Pi}
\def\ssw{{\sss W}}

\def\ssz{{\sss Z}}
\def\mw{M_{\sss W}}
\def\mz{M_{\sss Z}}
\def\smz{m_{\sss Z}}
\def\gf{G_{\sss F}}
\def\rht{{\sss R}}
\def\lft{{\sss L}}
\def\sm{{\sss SM}}
\def\ww{{\sss WW}}
\def\gg{{\gamma\gamma}}
\def\zz{{\sss ZZ}}
\def\zg{{{\sss Z}\gamma}}

\def\twe{\tw{e}}
\def\twm{\tw{m}_\ssz}
\def\tws{\tw{s}_w}
\def\twc{\tw{c}_w}
\def\sw{s_w}
\def\cw{c_w}
\def\gl{g_\lft}
\def\gr{g_\rht}
\def\gwk{$SU_\lft(2) \times U_{\sss Y}(1)$}


\rightline{McGill-93/13}
\rightline{UdeM-LPN-TH-93-151}
\rightline{hepph- 9306267}
\rightline{June 1993.}
\vskip .2in

\title
\centerline{Beyond S, T and U}
\endtitle

\authors
\centerline{Ivan Maksymyk,${}^a$ C.P. Burgess${}^b$ and David London${}^a$}  
\vskip .15in
\centerline{\it ${}^a$ Laboratoire de Physique Nucl\'eaire, l'Universit\'e de
Montr\'eal} 
\centerline{\it C.P. 6128, Montr\'eal, Qu\'ebec, CANADA, H3C 3J7.}
\vskip .1in
\centerline{\it ${}^b$ Physics Department, McGill University}
\centerline{\it 3600 University St., Montr\'eal, Qu\'ebec, CANADA, H3A 2T8.}
\endauthors

\abstract
The contribution to precision electroweak measurements due to TeV physics
which couples primarily to the $W^\pm$ and $Z$ bosons may be parameterized
in terms of the three `oblique correction' parameters, $S$, $T$ and $U$. We
extend this parameterization to physics at much lower energies, $\gsim 100$
GeV, and show that in this more general case neutral-current experiments
are sensitive to only two additional parameters. A third new parameter
enters into the $W^\pm$ width. 
\endabstract


\section{Introduction}

The standard electroweak theory has recently come of age, with experiments
now probing its predictions with sufficient accuracy to test its radiative
corrections in some detail. Besides providing a detailed test of the model,
these precision measurements are also very useful for the constraints they
impose on any potential new physics that might exist at energies higher
than those that have been hitherto experimentally explored. 

A particularly interesting class of new physics that is constrained by
these measurements consists of models which satisfy the following three
criteria 
\ref\peskin{M.E. Peskin and T. Takeuchi, \prl{65}{90}{964}; \prd{46}{92}{381}.}
\peskin:
\topic{(1)}
The electroweak gauge group must be \gwk, with no new electroweak gauge
bosons apart from the photon, the $W^\pm$ and the $Z$.
\topic{(2)}
The couplings of the new physics to light fermions are suppressed compared
to its couplings to the gauge bosons. 
\topic{(3)}
The intrinsic scale, $M$, of the new physics is large in comparison with
$\mw$ and $\mz$.
\endtopic

These criteria are particularly interesting principally for two reasons.
First, their implications for low-energy observables may be completely
described by three parameters, denoted $S$, $T$ and $U$ in Ref.~\peskin.
This allows these models to be meaningfully constrained as a group by
fitting for these parameters once and for all using the presently-available
precision electroweak data 
\ref\marciano{W.J. Marciano and J.L. Rosner, \prl{65}{90}{2963}.}
\ref\kennedy{D.C. Kennedy and P. Langacker, \prl{65}{90}{2967}.}
\marciano, \kennedy. Second, they include a large class of well-motivated
theories, such as technicolour models, models with extra generations,
multi-Higgs models, \etc

It is the purpose of this note to extend this analysis to theories which
satisfy the first two of the above criteria, but not the third. Since this
third item is used to neglect corrections to precision measurements that
are $O(\mz^2/M^2)$, there are two situations for which this type of
extension is necessary. The most obvious case is where the new physics is
comparatively light. Depending on the properties of the hypothetical new
particles, they could have masses as light as $M \lsim 100$ GeV and yet
still have escaped direct detection at LEP or the Tevatron. In this case
they are best constrained through their contributions through loops to
precision electroweak measurements, and we present here a formalism which
may be used to do so. 

The second type of application is to the scenario where new particle masses
are comparatively large, but where low-energy measurements become 
sufficiently accurate that $\mz^2/M^2$ corrections are no longer
negligible. This is the case that must be considered when using loop
effects to constrain anomalous three-gauge boson interactions, as has been
done in
\ref\tgvrefslin{K. Hagiwara, S. Ishihara, R. Szalapski and D. Zeppenfeld,
preprint MAD/PH/737, UT-635, KEK-TH-356, KEK preprint 92-214 (unpublished);
A. de R\'ujula, M.B. Gavela, P. Hernandez and E. Mass\'o, \npb{384}{92}{3};
P. Hern\'andez and F.J. Vegas, preprint CERN-TH 6670, LPTHE-Orsay 92/56,
FTUAM-92/34 (unpublished).}
\ref\sinha{D. Choudhury, P. Roy and R. Sinha, preprint TIFR-TH/93-08
(unpublished).} 
Refs.~\tgvrefslin\ and \sinha. As we show in more detail elsewhere
\ref\wwvpaper{C.P. Burgess, S. Godfrey, H. K\"onig, D. London and I.
Maksymyk, preprint McGill-93/14, OCIP/C-93-7, UdeM-LPN-TH-93-154
(unpublished).} 
\wwvpaper, the small size, $O(\mw^2/M^2)$, that is to be expected  for
these couplings implies that they can only be detected in precision
experiments to the extent that $O(\mw^2/M^2)$ cannot be neglected. In this
case the usual analysis in terms of $S$, $T$ and $U$ \sinha\ does {\it not}
apply, requiring our more general procedure. 

We show that the advantages of the previous formalism survive, and that the
implications of any such model for neutral-current data may in practice be
parameterized in terms of {\it four} parameters: Peskin and Takeuchi's $S$
and $T$, as well as two additional ones, which we call $V$ and $X$. If the
mass, width and low-energy couplings of the $W^\pm$ boson are also
included, only {\it two} more parameters are required: the quantity $U$ of
Ref.~\peskin, and one new variable, $W$. For practical applications $W$
often need not be considered, since it only arises in absolute measurements
of the $W^\pm$ widths, but cancels in its branching ratios.

Although this economy in the number of parameters required to parameterize
the data is similar to the economy that was found for physics at very high
scales, its origins here are very different. For physics at very high
scales only three parameters are possible {\it a-priori}, since in this
case the large value for $M$ permits an effective-lagrangian description in
which only the first few lowest-dimension effective interactions need be
considered. The same is not true in the present case, where the new degrees
of freedom can be comparatively light. The relatively few parameters which
do arise in this case reflect the fact that, at present, precision
measurements are made at only a very few scales, $q^2 \approx 0$ and $q^2 =
\mz^2$ or $\mw^2$. As a result the number of independent probes of new
physics is limited by the few scales at which this physics is sampled with
sufficient precision. This will certainly change in the future, such as at
LEP-200, once other scales become available for more detailed scrutiny. 

Our results reduce to the previous analyses in the limit that the new
physics is heavy. Since our treatment applies to both large and
comparatively small values for $M$, we are able to more quantitatively
identify the boundaries of applicability of the previous description. We do
so here by explicitly working through an example, in which we take the new
physics to consist of a degenerate doublet of new heavy fermions. We also
present this example as a sample of the type of diagnostic calculation that
will become necessary should a deviation from standard physics ever be
detected in the future, using these precision experiments. In this happy
event, a comparison between the sizes of the new parameters $V$, $W$ and
$X$ relative to the size of $S$ and $T$ can be used to infer the mass scale
that is associated with the underlying new physics.

\section{`Oblique' Corrections}

We start with the observation that, at present, precision electroweak
measurements exclusively involve the two-particle scattering of light
fermions. Given that the new physics is too heavy to be directly produced
in these experiments, there are three ways for it to indirectly contribute.
It can contribute to: ($a$) the propagation of the gauge bosons that can be
exchanged by the fermions, ($b$) the three-point fermion -- boson
couplings, and ($c$) the four-point direct fermion -- fermion interactions
(or `box'-diagram corrections). 

The importance of criteria {\it (1)} and {\it (2)} above is that when these
are satisfied then only corrections of type ($a$) --- the so-called
`oblique' corrections --- are dominant. (More precisely, we may use the
freedom to perform field redefinitions to put all of the new physics into
the vacuum polarizations. In fact, any new gauge-fermion or four-fermion
vertex in which the fermions appear only through linear combinations of the
total SM currents may be re-expressed in this way. This freedom to recast
the effective lagrangian is exploited in Ref.~\wwvpaper.) Under these
circumstances the complete impact of any new heavy degrees of freedom
arises through their contributions to the gauge boson vacuum polarizations,
$\Pi^{\mu\nu}_{ab}(q) = \Pi_{ab}(q^2) g^{\mu\nu} + (q^\mu q^\nu$ terms),
with $a,b = \gamma,W^\pm,Z$. We therefore supplement the standard model
(SM) by adding a new-physics contribution to the gauge boson vacuum
polarizations:
\eq
\Pi_{ab}(q^2) = \Pi^\sm_{ab}(q^2) + \dpi_{ab}(q^2).
\eeq
The first contribution represents the SM contribution, including all
appropriate radiative corrections, while all new-physics effects are
contained in the second term. 

\ref\grinwise{For an analysis which goes to next order in the $q^2$
expansion, see B. Grinstein and M.B. Wise, \plb{265}{91}{326}.}
In Refs.~\peskin-\kennedy, it was assumed that the $\dpi_{ab}(q^2)$ are due
to new physics at a very high mass scale, and are thus well-described by a
Taylor expansion to linear order in $q^2$: $\dpi_{ab}(q^2) \approx A_{ab} +
B_{ab} q^2$ \grinwise. Under this assumption it is straightforward to
determine the number of independent parameters required to describe all
new-physics effects. The reasoning goes as follows. There are eight
quantities describing the new physics: $A_{ab} = \dpi_{ab}(0)$ and $B_{ab}
= \dpi_{ab}'(0),$, with the pair $(ab)$ taking the four independent values:
$(ab) = (\gamma\gamma)$,  $(Z\gamma)$, $(ZZ)$ and $(WW)$. (The prime
denotes differentiation with respect to $q^2$: $\dpi' \equiv d \,\dpi /
dq^2$.) Two of these---$\dpi_\gg(0)$ and $\dpi_\zg(0)$---are automatically
zero by gauge invariance. Three other linear combinations can be eliminated
when the three input parameters---say, $\alpha$, $\mz$ and $\gf$---are
renormalized. Thus, all new physics effects can be described by  three
combinations of the $\dpi$'s, denoted $S$, $T$ and $U$ in Ref.~\peskin\
(the precise definitions of these parameters are given later).

In the more general case we cannot assume any specific form for
$\dpi_{ab}(q^2)$, and so there are potentially many more parameters that
can enter into physical observables. Our purpose in this section is to
compute the dependence of well-measured quantities on these oblique
corrections, and to show that in fact only a few additional parameters
arise beyond the three that were introduced by Peskin and Takeuchi. This
remarkable simplification reflects the fact that current precision
measurements sample the vacuum polarizations only at a few values of $q^2$:
$q^2 \approx 0$ and $q^2 = \mw^2$ or $\mz^2$. 

We first compute the experimental implications of the $\dpi_{ab}(q^2)$. 
This task is made much simpler by the very success of the standard model
itself. The agreement between the data and the SM predictions, including
radiative corrections, implies that the $\dpi_{ab}(q^2)$ cannot be larger
than at most $O(1\%)$ of the size of their tree-level SM counterparts. It
follows that we may simply perturb in the oblique corrections --- with one
important exception which we treat in detail below --- and stop at linear
order. Furthermore, SM radiative corrections to any new-physics
contribution may also be ignored to within the accuracy we require. We may
therefore simply work to tree-level in the oblique corrections,
$\dpi_{ab}$, and then simply add the result to the corresponding SM
contribution, including potential radiative corrections. 

\subsection{Shifting the Standard-Model Couplings}

There are two distinct ways in which these quantities can enter into
predictions for any particular observable. Besides contributing directly to
predictions for the quantities of interest, they can also change the
numerical values that are inferred from experiment for the various SM
parameters, such as for the electric charge, $e$, $\sw \equiv
\sin\theta_w$, \etc\ This change then shifts the SM prediction for all
other quantities. We first compute this shift.

The standard electroweak interactions are parameterized by three variables,
(in addition to other parameters, like fermion masses, which do not concern
us here) which we denote as $\twe$, $\tws$, and $\twm$. The values for
these are fixed by comparing SM predictions to the three best-measured
observables, which we take to be: ($i$) the fine-structure constant,
$\alpha$, as measured in low-energy electron scattering, Fermi's constant,
($ii$) $\gf$, as measured in muon decay, and ($iii$) the $Z$ boson mass,
$\mz$, as measured at LEP. In order to calculate the change that the new
physics implies for these parameters we must compute the contribution of
the $\dpi_{ab}$'s to these quantities. 

Following the above reasoning, we compute the corrections to $\alpha$ and
$\gf$ by computing the corrections to low-energy electron-scattering and to
muon decay, working to tree level in the oblique corrections. The shift in
the $Z$ boson mass follows from its definition in terms of the pole of the
propagator. This leads to the following expressions:  
\label\shifts
\eq \eqalign{
\alpha &= \alpha_\sm(\twe) \Bigl[ 1 + \dpi_\gg'(0) \Bigr], \cr
\gf &= (\gf)_\sm(\twe,\tws,\twm) \left[ 1 - { \dpi_\ww(0) \over \mw^2}
\,\right], \cr
\mz^2 &= (\mz^2)_\sm(\twe,\tws,\twm) \left[ 1 + { \dpi_\zz(\mz^2) \over \mz^2}
\, \right], \cr} \eeq
We may define `standard' parameters by equating the left-hand sides of
these expressions to their SM formulae. For example, we define $e$ by
requiring $\alpha = \alpha_\sm(e) \equiv 4\pi e^2 + $(loops), $\gf =
e^2/(4\sqrt{2} \sw^2 \cw^2 \smz^2) + $(loops), \etc. This leads to the
following expressions: 
\label\parameters
\eq \eqalign{
\twe &= e \left[ 1 - \hf \; \dpi'_\gg(0) \,\right] ,\cr
\tws^2 &= \sw^2 \left[ 1 - { \cw^2 \over \cw^2 - \sw^2} \left( \dpi_\gg'(0) 
- {\dpi_\zz(\mz^2) \over \mz^2} + {\dpi_\ww(0) \over \mw^2} \right) \right],\cr
\twm^2 &= \smz^2 \left[ 1 - { \dpi_\zz(\mz^2) \over \mz^2 } \,\right]. \cr} \eeq
In our notation $\smz$ denotes the standard model parameter, as opposed to
the physical quantity, $\mz$. We make this distinction, since these can
differ once SM radiative corrections are included.

The prediction for any other observable, $A$, may now be written $A = 
A_\sm(\twe,\tws,\twm) + \delta A$, where the first term is the SM
prediction, and where the second term is the `direct' contribution of the
new oblique corrections to the observable in question. In order to take
advantage of the most precise radiatively corrected SM calculations, it is
then useful to re-express $A$ using eqs.~\parameters, as $A =
A_\sm(e,\sw,\smz) + \delta A'$, where $A_\sm(e, \sw, \smz)$ now takes the
same numerical value as it does in the standard model in the absence of new
physics. 

\subsection{Low-Energy Observables: $S$, $T$ and $U$}

The direct contributions to observables from the new physics are now easily
computed. The first step comes in choosing which observables to compute.
Here we meet a further simplification. The main point is that most, if not
all, {\it precision} electroweak measurements are performed at very few
scales. They are  either at very low energies, $q^2 \ll \mw^2$, or at the
$Z$ resonance. To the extent that the $W^\pm$ mass and widths are also
regarded as being sufficiently precisely measured, then $q^2 = \mw^2$ is
also relevant. This implies that the vacuum polarizations are only sampled
at these very few scales. As a result, even though $\dpi_{ab}(q^2)$ can
potentially depend on $q^2$ in a complicated way, its implications for
precision measurements can be summarized in a few numbers. 

We concentrate first on low-energy observables, for which we may take $q^2
\approx 0$. (We defer the treatment of quantities that are defined near the
mass shell, $q^2 = \mw^2$ and $\mz^2$ until later, since for these there is
an added complication.) Examples of observables of this type include the 
low-energy electron scattering and muon decay processes of the previous
section, as well as deep-inelastic neutrino scattering, atomic parity
violation experiments \etc\ In this case, the new physics contributions may
be straightforwardly worked out perturbatively in the $\dpi_{ab}$. 

For example, low-energy measurement of parity-violating asymmetries, such
as $A_{\sss LR}$ and $A_{\sss FB}$ in electron scattering, can be used to
define an effective value for $\sin^2 \theta_w$. This is given by:
\label\lowenergys
\eq \eqalign{
(\sw^2)_{\rm eff}(q^2 \approx 0) &= \tws^2 - \sw \cw \; \dpi_\zg'(0) \cr
&= \sw^2 \left[ 1 - {\cw^2 \over \cw^2 - \sw^2} \left( \dpi_\gg' -
{\dpi_\zz(\mz^2) \over \mz^2} + {\dpi_\ww(0) \over \mw^2} \right) - {\cw
\over \sw} \dpi'_\zg(0) \right]. \cr} \eeq

Clearly the effects of the new physics on any such observable are given by
the induced shifts in the SM parameters, as well as a linear combination of
the various $\dpi_{ab}$'s --- or their derivatives --- evaluated at $q^2
\approx 0$. As a result they never probe the oblique corrections beyond
linear order in their expansions in powers of $q^2$, and so they are
completely described in terms of the three Peskin-Takeuchi parameters, $S$,
$T$ and $U$: 
\label\stu
\eqa 
{\alpha S \over 4 \sw^2 \cw^2 } &= \left[\, {\dpi_\zz(\mz^2) - \dpi_\zz(0)
\over \mz^2} \,\right] - {(\cw^2 - \sw^2) \over \sw\cw} \; \dpi_\zg'(0) -
\dpi'_\gg(0), \eol   
\alpha T &= {\dpi_\ww(0) \over \mw^2} - {\dpi_\zz(0) \over \mz^2}, \eol  
{\alpha U \over 4 \sw^2 } &= \left[\, {\dpi_\ww(\mw^2) - \dpi_\ww(0) \over
\mw^2} \, \right] - \cw^2 \left[ \, { \dpi_\zz(\mz^2) - \dpi_\zz(0) \over \mz^2}
\, \right] \eolnn
&\qquad  -  \sw^2 \dpi_\gg'(0) - 2 \sw \cw \dpi_\zg'(0).\eeol  
\eeq  
These definitions are deliberately cast in a way that does not assume that
the $\dpi_{ab}(q^2)$ are linear functions of $q^2$. Notice that although
these expressions agree with the formulation of $S$, $T$ and $U$ given in
Ref.~\marciano, the analysis of these authors only applies when
$\dpi_{ab}(q^2) = A_{ab} + B_{ab}q^2$. With these definitions
eq.~\lowenergys\ takes the usual form:
\label\usualform
\eq
{(\sw^2)_{\rm eff}(q^2 \approx 0)  \over (\sw^2)_\sm} = 
1 + {\alpha S \over 4 \sw^2 ( \cw^2 - \sw^2)} 
- { \cw^2 \; \alpha T \over \cw^2 - \sw^2}~.
\eeq
Any other low-energy observable may be analyzed in a similar fashion.

\subsection{Observables on Mass Shell: Beyond $S$, $T$ and $U$}

A slightly different analysis is required when considering observables that
are defined at $q^2 = \mz^2$ or $\mw^2$, as is appropriate for the masses
and widths of the gauge bosons themselves. This is because it is no longer
a good approximation to simply work to linear order in the two quantities
$\dpi_\zz(\mz^2)$ and $\dpi_\ww(\mw^2)$ when these quantities are inserted
into boson lines which are on shell. This is because when these insertions
are combined with the very small denominators of the corresponding gauge
boson propagators in these lines (which are themselves $O(\dpi)$) their
contributions can be $O(1)$. In this case the vacuum polarization
insertions must be summed to all orders in the usual way, using the
Schwinger-Dyson equations. 

The result is simple to state, however. The shift in the gauge boson mass
is simply given by the (real part of the) position of the pole in the
corresponding propagator. We take as our first example the mass of the
$W^\pm$, which is given in the SM by $(\mw^2)_\sm(\twe,\tws,\twm) = \twm^2
\twc^2 + $(loop corrections). The oblique corrections change this to 
\label\firstwmass
\eq 
\mw^2 =  (\mw^2)_\sm(\twe,\tws,\twm) \left[ 1 + { \dpi_\ww(\mw^2) \over \mw^2}
\, \right]. \eeq
Shifting to the `standard' parameters then gives:
\label\wmassforreal
\eq \eqalign{
\mw^2 &=  (\mw^2)_\sm(e,\sw,\smz) \left[ 1 - {\cw^2 \over \cw^2 - \sw^2} \; 
{\dpi_\zz(\mz^2) \over \mz^2}\right.  \cr
& \qquad \left. + {\sw^2 \over \cw^2 - \sw^2} \left( \dpi'_\gg(0)
+ {\dpi_\ww(0) \over \mw^2} \right) + {\dpi_\ww(\mw^2) \over \mw^2}\right]. \cr
&=  (\mw^2)_\sm(e,\sw,\smz) \left[ 1 - {\alpha S \over 2(\cw^2 - \sw^2)}
+ {\cw^2 \alpha T \over (\cw^2 - \sw^2)} + {\alpha U \over 4 \sw^2} \right]. \cr}
\eeq

Let us now consider the $W^\pm$ width. This is obtained by multiplying the
lowest-order SM result by the renormalization factor, $\Scz_\ssw = 1 +
\dpi'_\ww(\mw^2)$, that arises due  to the use of the fully summed
propagator. As a result:
\eq
\delta \Gamma(W \to e\nu) = {\twe^2 \mw \over 48 \pi \tws^2} \;
\dpi'_\ww(\mw^2). 
\eeq
Notice that we use the full mass, $\mw$, in this expression since this is
what appears in the phase space integration. Transforming to `standard'
parameters leads to:
\eq\eqalign{
{\Gamma(W\to \hbox{all}) \over \Gamma_\sm(W\to \hbox{all}) }
&= {\Gamma(W\to e\nu) \over \Gamma_\sm(W\to e\nu) } \cr
&=  1 + {1 \over \cw^2 - \sw^2} \left[ \sw^2\, \dpi'_\gg(0) + \cw^2  
\left( {\dpi_\ww(0) \over \mw^2} - {\dpi_\zz(\mz^2) \over \mz^2} \right)
\right] + \dpi'_\ww(\mw^2). \cr} \eeq
The $Z$ width into neutrinos may be computed in an identical way. 
Proceeding along precisely the same lines as for the $W^\pm$ width, and
using $\Scz_\ssz = 1 + \dpi_\zz'(\mz^2)$, gives the following result:
\eq
{\Gamma(Z \to \nu\ol{\nu}) \over \Gamma_\sm(Z \to \nu\ol{\nu}) } 
=  1 - {\dpi_\zz(\mz^2) \over \mz^2} +  {\dpi_\ww(0) \over \mw^2}
+ \dpi'_\zz(\mz^2).  \eeq

A new feature enters into the calculation of the $Z$-boson widths into
charged-particle final states, since these receive contributions from 
$\dpi_\zg(\mz^2)$. Unlike the effects due to the insertion of $\dpi_\zz$,
the contribution of this term may be computed perturbatively, since the
gauge boson line is not forced to be on shell. The result is a contribution
to the effective value for $\sw^2$ that is measured in $A_{\sss LR}$,
$A_{\sss FB}$ at the $Z$ resonance. We find:
\eq 
{(\sw^2)_{\rm eff}(q^2 \approx  \mz^2) \over (\sw^2)_\sm} =  {(\sw^2)_{\rm eff}
(q^2 \approx  0) \over (\sw^2)_\sm}  - \sw \cw \; \left[ { \dpi_\zg(\mz^2) 
\over \mz^2} -  \dpi_\zg'(0) \right].  \eeq

At this point a dramatic simplification occurs. All of the new physics then 
combines into compact expressions involving Peskin and Takeuchi's three
parameters, as well as the following three new ones, which we call $V$, $W$
and $X$:
\label\vwx
\eqa
\alpha V &= \dpi_\zz'(\mz^2) - \left[ \, {\dpi_\zz(\mz^2) - \dpi_\zz(0)
\over \mz^2}  \, \right], \eol 
\alpha W &= \dpi_\ww'(\mw^2) - \left[ \, {\dpi_\ww(\mw^2) - \dpi_\ww(0) \over
\mw^2}  \, \right]  \eol 
\alpha X &= - \sw \cw \left[ \, { \dpi_\zg(\mz^2) \over \mz^2} - \dpi'_\zg(0)
 \, \right]. \eeol 
\eeq
Notice that these expressions would vanish if $\dpi_{ab}(q^2)$ were simply
a linear function of $q^2$. 

In terms of these parameters our previous expressions for the $W^\pm$ and
$Z$ widths, and $(\sw^2)_{\rm eff}(\mz^2)$, become:
\eqa
{\Gamma(W\to \hbox{all}) \over \Gamma_\sm(W\to \hbox{all}) }
&= 1 - { \alpha S \over 2 (\cw^2 - \sw^2)} 
+ { \cw^2\; \alpha T \over (\cw^2 - \sw^2)} 
+ {\alpha U \over 4\sw^2} + \alpha W~, \eol
{\Gamma(Z \to \nu\ol{\nu}) \over \Gamma_\sm(Z \to \nu\ol{\nu}) } 
&=   1 + \alpha T + \alpha V, \eol
{(\sw^2)_{\rm eff}(q^2 \approx  \mz^2) \over (\sw^2)_\sm} 
&= 1 + {\alpha S \over 4 \sw^2 ( \cw^2 - \sw^2)} 
- { \cw^2 \; \alpha T \over \cw^2 - \sw^2} + \alpha X~. \eeol
 \eeq
Notice that one of these new parameters, $W$, turns out to appear only in
this expression for $\Gamma_\ssw$. The $Z$ width into any fermion pairs may
be expressed in terms of the remaining two parameters, $V$ and $X$, where
$V$ describes a contribution to the overall normalization of the strength
of the interaction, and $X$ acts to shift the effective value of
$(\sw^2)_{\rm eff}$. 

\subsection{Numerical Results}

We tabulate the contributions to some precision measurements in Table (1).
In preparing this table we use the following numerical values in obtaining
these results: $\alpha(\mz) = 1/127.8$, $\sw^2(\mz) = 0.2323$, and $\mz =
91.17$ GeV. For the SM predictions we choose the fiducial values, $m_t =
140$ GeV and $m_{\sss H} = 100$ GeV. When appropriate our numbers clearly
reduce to the results of Ref.~\marciano, from whom we also have taken the
experimental limits.

Several features come to light on inspection of Table (1). First, for
neutral-current data at low energies and on the $Z$ resonance, only the
four parameters $S$, $T$, $V$ and $X$ arise.\foot\renorm{The absence of
{\it U} in these expressions follows from our choice of three 
standard-model input observables.} Of these, only the standard two, $S$ and
$T$, contribute to low-energy observables for which $q^2 \approx 0$, since
$V$ and $X$ appear only in observables that are defined at $q^2 = \mz^2$.
As a result all of the predictions for low-energy quantities agree with
earlier work \marciano. In particular, the favouring of negative values for
$S$ by the cesium atomic parity violation experiments is not affected by
the introduction of the additional parameters. 

Next, although the $Z$ results are the most precise, some $W^\pm$
properties, such as measurements of $\mw$, are sufficiently accurate to
competitively bound the relative parameters. For these charged-current
observables two more parameters enter: the usual quantity $U$ for
quantities defined at $q^2 \simeq 0$ (as well as $\mw$), and the parameter
$W$ in the $W^\pm$-boson decay widths.

We next apply these results to an illustrative example.

\section{`Technifermions': An Example}

New, massive fermions which carry electroweak quantum numbers, but which do
not mix appreciably with ordinary light fermions, furnish an concrete model
to which the above reasoning applies. If these fermions are sufficiently
heavy, say $m \gsim 1$ TeV, their implications may be summarized as
contributions to $S$, $T$ and $U$. We wish to explore here the much lighter
mass range, $m \sim$ (several hundred GeV), that can still be consistent
with such particles not having been detected at current accelerators. Our
goal is to show how the formalism just presented can be used to constrain
the properties of such particles. As a bonus we can vary the fermion mass,
and determine quantitatively at what point the usual three-parameter
description becomes sufficiently accurate. 

We therefore require the vacuum polarization that is induced by such a
collection of fermions. Evaluating the graph of Fig.~(1) produces the
following result: 
\label\vacpol
\eq
\dpi_{ab}(q^2) = {1 \over 2 \pi^2} \sum_{ij} \int_0^1 dx \; f_{ab}(q^2,x)
\; \ln \left[ { m^2_{ij}(x) - q^2 x (1-x) \over \mu^2 } \right] , \eeq
where $m^2_{ij}(x) \equiv m^2_i (1-x) + m^2_j x$, and
\eq
f_{ab}(q^2,x) = {\gl^a \gl^{b*} + \gr^a \gr^{b*} \over 2} \left[ x (1-x) \,
q^2 - { m^2_{ij}(x) \over 2} \right] + {\gl^a \gr^{b*} + \gr^a \gl^{b*} \over
2} \; \left( { m_i m_j \over 2} \right). \eeq
In these expressions, $m_i$ and $m_j$ are the masses of the fermions which
circulate in the loop, and $\gl^a$ and $\gr^a$ represent their left- and
right-handed couplings to the gauge-bosons: $a = \gamma, W^\pm, Z$. For a
standard-model doublet $\gl^\gamma = \gr^\gamma = e Q_i$, $\gl^\ssz =
(e/\sw\cw) [T_{3i} - Q_i \sw^2]$,  $\gr^\ssz = (e/\sw\cw) [ - Q_i \sw^2]$, 
$\gl^\ssw = e / \sqrt{2} \sw$ and $\gr^\ssw = 0$. We have renormalized
$\dpi_{ab}$ using $\ol{\hbox{MS}}$, and $\mu^2$ is the associated
renormalization scale. 

For simplicity we consider only the specific case of one additional doublet
of degenerate leptons, for which $m_i = m_j \equiv m$. Then
\label\vacpolresults
\eqa
\dpi_\gg (q^2) &= {e^2 q^2 \over 2 \pi^2} \int_0^1 dx \; 
x (1-x) \ln \left[ { m^2 -q^2 x(1-x) \over \mu^2} \right], \eol
\dpi_\zg (q^2) &= {e^2 \sw q^2 \over 2 \pi^2 \cw} \int_0^1 dx \; 
x (1-x) \ln \left[ { m^2 -q^2 x (1-x) \over \mu^2} \right], \eol
\dpi_\ww (q^2) &= {e^2 \over 8 \pi^2 \sw^2} \int_0^1 dx \; 
\left( q^2 x (1-x) - {m^2 \over 2} \right) \; 
\ln \left[ { m^2  -q^2 x (1-x) \over \mu^2} \right], \eol 
\dpi_\zz (q^2) &= {e^2 \over 16 \pi^2 \sw^2 \cw^2} \int_0^1 dx \; 
\Bigl[ (1 - 2\sw^2 + 4 \sw^4) q^2 x (1-x) - m^2 \Bigr] \; 
\ln \left[ { m^2 -q^2 x (1-x) \over \mu^2} \right]. \eolnn
& \eeol \eeq
Using these expressions in the definitions, eqs.~\stu\ and \vwx, gives the
parameters $S$ through $X$ as functions of the mass of the doublet. The
parameter $T$, which is a measure of custodial symmetry breaking, vanishes
since the doublet is degenerate. We plot the behaviour of the remainder of
these parameters against $m$ in Fig.~(2). We use $\mu = m$ in this figure.

\ref\takref{We thank T. Takeuchi for correspondence on this point.} 
The curves in Fig.~(2) verify the dominance of the parameter $S$ when $m$
is large. The parameters $V$, $W$ and $X$ fall quite quickly to zero, as
might be expected. Interestingly, they are always larger than $U$, even for
$m$ as large as 1 TeV. This is because in this particular example the
custodial symmetry is unbroken \takref. In general, however, $U$ is not
expected to be quite this small. Nevertheless, this does suggest that if
fits which include $U$ are to be performed using precision data, then the
same fits should also include $V$, $W$ and $X$.

\section{Conclusions}

The $S$--$T$--$U$ parameterization of oblique electroweak corrections, as
presented in Ref. \peskin, has proven to be a useful tool for constraining
new physics from above the electroweak scale. It is useful because it
summarizes into a few quantities the implications for precision electroweak
experiments of a broad class of interesting models. One of its limitations
is that it can only be applied when the scale, $M$, of new physics is high
enough to justify the neglect powers of $\mz^2/M^2$. We have extended the
analysis to the case where the threshhold for new physics is too low to
justify this approximation. 

We find that even for the case of comparatively light new physics,
precision electroweak measurements are only sensitive to a small number of
independent parameters. Precisely three new ones are required, which we
call $V$, $W$ and $X$. The dependence on these parameters of many of the
well-measured observables is summarized in Table (1). Neutral-current data
is completely described by four quantities, $S$, $T$, $V$ and $X$, of which
the latter two only contribute to observables defined at the $Z$ resonance.
A description of $W^\pm$ physics requires both the Peskin-Takeuchi $U$
parameter, and the additional variable $W$, although $W$ will only be
relevant once measurements of the $W^\pm$ width considerably improve. 

So few parameters suffice because at present precision experiments are 
confined to light-fermion scattering with four-momentum transfer that is
either equal to, or much lower than, $\mw$ or $\mz$. As a result the
$q^2$-dependence of any oblique corrections is only sampled for two or
three places. This emphasizes the importance of performing precision
measurements at other values of $q^2$, such as will be possible at LEP-200,
since these experiments can probe complementary facets of the underlying
physics. 

We have illustrated how the parameters $S$ through $X$ can arise using a
particularly simple example of a degenerate doublet of heavy fermions. This
permits us to quantitatively follow the heavy-mass dependence of all six
quantities. We verify how $S$ comes to dominate as the doublet mass grows.
In our example, however, the parameter $U$ is never larger than $V$, $W$
and $X$. In the future such an analysis could prove useful if a discrepancy
between the standard model and these measurements were ever to arise. In
this case the comparison between the sizes of the new parameters with $S$
and $T$ can be used to infer the mass scale of the new physics that is
involved. 

\bigskip
\centerline{\bf Acknowledgments}
\bigskip

We thank T. Takeuchi for helpful comments. This research was partially
funded by funds from the N.S.E.R.C.\ of Canada and les Fonds F.C.A.R.\ du
Qu\'ebec. 

\vfill\eject
\centerline{\bf Figure Captions}
\bigskip

\topic{Figure  (1)}
The Feynman graph through which the heavy fermion doublet contributes
to the gauge boson vacuum polarization. 

\topic{Figure (2)}
A plot of the oblique-correction parameters $S$, $U$, $V$, $W$ and $X$ against 
the mass of the heavy doublet of degenerate fermions which generate them. 
The parameter $T$ vanishes identically because the doublet is degenerate. 

\listrefs

\bye